# Computing Eigen-Emittances from Tracking Data*

Y. Alexahin (FNAL APC)

## 1. Introduction

Finding normal mode (eigen) emittances from experimental or simulations data is needed for many applications, most notably in analysis of particle cooling.

In the second section we show – using the theory developed by V. Lebedev & A.Bogacz [1] – how to find eigen-emittances and optics functions from a known covariance matrix in the case of coupled oscillations. As an application of the developed formalism we make an estimate of the error in eigen-emittances if the mechanical momenta are used instead of the canonical ones.

The third, fourth and fifth sections are devoted to algorithms for suppression of the halo contribution to the covariance matrix. First, we show that simple introduction of weight depending on the particle position in the phase space does not allow for the necessary precision in absence of halo.

In the fourth section we describe an algorithm of nonlinear fitting of an ensemble of particles in the phase space with a distribution function of specified type (e.g. Gaussian) which reduces the contribution of "outliers". In the fifth section this algorithm is reduced to a system of equations which can be solved iteratively. It is important to note that the found equation for covariance matrix cannot be obtained by introducing a weight function.

In the last, sixth section we present an example of application of the developed algorithm to the 6D ionization cooling of muons.

## 2. Computation of Eigen-Emittances from Covariance Matrix

First let us introduce notation conventions: underlined characters will denote (column) phase space vectors, whereas upright capital letters will be used to designate matrices.

In our treatment of coupled betatron motion we closely follow Ref.[1] making some minor changes in notations and extending it to the 6D case. Let us recall the basic relations for the sake of completeness.

Following [2] let us choose the path length $s$ along the reference orbit as the independent variable and dynamical variables in the form

$$\underline{z} = \{x, P_x, y, P_y, s - c\beta_0 t, \delta\}, \tag{1}$$

where $P_{x,y}$ are canonical momenta normalized by the reference value $p_0 = mc\beta_0\gamma_0$:

$$P_x = (p_x + \frac{e}{c} A_x) / p_0 \tag{2}$$

with $p_{x,y}$ and $A_{x,y}$ being the components of the mechanical momentum and magnetic vector potential respectively (we use Gaussian units). Finally,

$$\delta = (\gamma - \gamma_0) / \beta_0^2 \gamma_0. \tag{3}$$

Suppose that from measurements or simulations we have a set of $N$ particles positions in the phase space, $\underline{z}^{(k)}$, $k = 1,\ldots, N$, and our task is to find the normal mode emittances. Let us – for the length of this section – assume that the given distribution does not contain long tails (outliers) and therefore we can use simple averaging to find elements of the covariance matrix $\Sigma$:

$$\Sigma_{i,j} = \frac{1}{N} \sum_{k=1}^{N} \zeta_i^{(k)} \zeta_j^{(k)}, \quad \underline{\zeta}^{(k)} = \underline{z}^{(k)} - \underline{a}, \quad \underline{a} = \frac{1}{N} \sum_{k=1}^{N} \underline{z}^{(k)} \tag{4}$$

If particle have weights $w_k$ – like in simulations with G4beamline [3] or other codes – the average values should be taken as

$$\frac{1}{N}\sum_{k=1}^{N} q^{(k)} \to \sum_{k=1}^{N} q^{(k)} w_k \bigg/ \sum_{k=1}^{N} w_k \qquad (5)$$

---

*) Work supported by Fermi Research Alliance, LLC under Contract DE-AC02-07CH11359 with the U.S. DOE.

Now we make our basic assumption that the particle distribution in the phase space is a function of the quadratic form (or, more precisely, of half it):

$$\Phi(\underline{\zeta}) = (\underline{\zeta}, \Sigma^{-1}\underline{\zeta}) \equiv \sum_{i=1}^{6} \zeta_i (\Sigma^{-1}\underline{\zeta})_i = \sum_{i,j=1}^{6} \Sigma_{ij}^{-1} \zeta_i \zeta_j . \qquad (6)$$

The $\Sigma$-matrix – and its inverse as well – are symmetric and positive-definite by construction and therefore have real positive eigenvalues. Using its eigenvectors as columns of matrix V we can construct transformation

$$\underline{\zeta} = V\underline{\xi} . \qquad (7)$$

which reduces $\Phi$ to a sum of squares of new variables divided by the corresponding eigenvalues. But generally neither these variables nor the eigenvalues have an immediate physical meaning unless transformation (7) is canonical. For that its matrix must satisfy the symplecticity condition:

$$V^t S V = S, \qquad (8)$$

where S is a symplectic unity matrix and superscript "t" means transposition. With the chosen order of variables in Eq.(1) S has the form

$$S = \begin{pmatrix} 0 & 1 & 0 & 0 & 0 & 0 \\ -1 & 0 & 0 & 0 & 0 & 0 \\ 0 & 0 & 0 & 1 & 0 & 0 \\ 0 & 0 & -1 & 0 & 0 & 0 \\ 0 & 0 & 0 & 0 & 0 & 1 \\ 0 & 0 & 0 & 0 & -1 & 0 \end{pmatrix}, \qquad (9)$$

It has the following properties: $S^t = -S$, $S^2 = -I$, where I is identity matrix, and det S=1.

<u>Eigensystem of the associate $\Sigma$-matrix.</u>

In Ref.[1] it was shown how to reduce $\Phi$ to a sum of squares when a canonical transformation to the normal form variables is known from the underlying Hamiltonian mechanics. Here we offer a solution of the inverse problem: constructing the required canonical transformation from the $\Sigma$-matrix itself without reference to the underlying Hamiltonian system. More than that, we show how information on the beta and dispersion functions can be recovered in the process.

Taking the cue from Ref.[1] let us consider the matrix product

$$\Omega = S\Sigma^{-1} . \qquad (10)$$

It has some important properties:

1. *If $\lambda$ is an eigenvalue of $\Omega$ then $-\lambda$ is also its eigenvalue.*

To prove this let us remind that eigenvalues are solutions of the equation

$$\det(\Omega - \lambda I) = 0 \qquad (11)$$

and that det $A^t$ = det A for any A. Therefore

$$\det(\Omega - \lambda\,\mathrm{I})^t = \det(\Omega - \lambda\,\mathrm{I}) = 0$$
$$= \det(\Omega^t - \lambda\,\mathrm{I}) = \det(-\Sigma^{-1}S - \lambda\,\mathrm{I}) = \det(-\Sigma^{-1} + \lambda\,S)\,S = \qquad(12)$$
$$= \det(S^2\Sigma^{-1} + \lambda\,S) = \det S(\Omega + \lambda\,\mathrm{I}) = \det(\Omega + \lambda\,\mathrm{I})$$

where we took into account the symmetry of $\Sigma^{-1}$.

2. *Squares of eigenvalues of $\Omega$ are real negative numbers.*

This follows directly from the fact that matrices $\Sigma^{-1}$ and $S^{-1}\Sigma^{-1}S$ are similar and have the same set of positive eigenvalues so that their product is also positive-definite. Thereby $\Omega^2$ is negative-definite:

$$\Omega^2 = S\Sigma^{-1}S\Sigma^{-1} = -(S^{-1}\Sigma^{-1}S)\Sigma^{-1}. \qquad(13)$$

As a consequence of the above properties the eigenvalues of $\Omega$ form purely imaginary conjugate pairs which we will order as

$$\lambda_{2m-1} = -\frac{i}{\varepsilon_m}, \quad \lambda_{2m} = \frac{i}{\varepsilon_m}, \qquad(14)$$

where $m=1, 2, 3$ is the normal mode number. In the following we assume all $\varepsilon_m$ to be different.

To prove that $\varepsilon_m$ really are the eigen-emittances let us start with the properties of the matrix $\Omega$ eigenvectors $\underline{v}_i$.

$$\lambda_j(\underline{v}_i, S\underline{v}_j) = (\underline{v}_i, S\Omega\underline{v}_j) = -(\underline{v}_i, \Sigma^{-1}\underline{v}_j) = -(\Sigma^{-1}\underline{v}_i, \underline{v}_j) = (S^2\Sigma^{-1}\underline{v}_i, \underline{v}_j) =$$
$$= (S\Omega\underline{v}_i, \underline{v}_j) = -(\Omega\underline{v}_i, S\underline{v}_j) = -\lambda_i(\underline{v}_i, S\underline{v}_j) \Rightarrow \qquad(15)$$
$$(\lambda_i + \lambda_j)(\underline{v}_i, S\underline{v}_j) = 0$$

where we again invoked the symmetry of $\Sigma^{-1}$. The last equation shows that eigenvectors belonging to different normal modes are skew-orthogonal. For eigenvectors belonging to the same normal mode we can put

$$\underline{v}_{2m} = \underline{v}_{2m-1}^* \qquad(16)$$

where asterisk denotes complex conjugation and impose the following normalization:

$$(\underline{v}_{2m-1}^*, S\underline{v}_{2m-1}) = -2i \qquad(17)$$

Taking real and imaginary parts of eq.(17) we can rewrite the orthonormality conditions in the form

$$(\underline{v}_i', S\underline{v}_j') = (\underline{v}_i'', S\underline{v}_j'') = 0,$$
$$(\underline{v}_{2m-1}', S\underline{v}_{2n-1}'') = \delta_{mn}, \qquad(18)$$
$$\underline{v}_i' \equiv \mathrm{Re}\,\underline{v}_i, \quad \underline{v}_i'' \equiv \mathrm{Im}\,\underline{v}_i$$

From (14) we have

$$\Omega\underline{v}_{2m-1}' = \frac{1}{\varepsilon_m}\underline{v}_{2m-1}'', \quad \Omega\underline{v}_{2m-1}'' = -\frac{1}{\varepsilon_m}\underline{v}_{2m-1}' \qquad(19)$$

<u>Transformation to normal form variables.</u>

Now, following Ref. [1], let us build a matrix taking real and (with a minus sign) imaginary parts of the eigenvectors as its columns

$$V = \{\underline{v}_1', -\underline{v}_1'', \underline{v}_3', -\underline{v}_3'', \underline{v}_5', -\underline{v}_5''\} \qquad (20)$$

Taking into account eqs.(18) and (19) it is easy to verify that this matrix satisfies symplecticity condition (8) and reduces $\Omega$ to the form

$$V^{-1} \Omega V = S \Xi,$$
$$\Xi = \mathrm{diag}(\frac{1}{\varepsilon_1}, \frac{1}{\varepsilon_1}, \frac{1}{\varepsilon_2}, \frac{1}{\varepsilon_2}, \frac{1}{\varepsilon_3}, \frac{1}{\varepsilon_3}). \qquad (21)$$

With the help of transformation (7) and condition (8) we can bring quadratic form (6) to the normal form

$$\Phi = (\underline{\zeta}, \Sigma^{-1} \underline{\zeta}) = (V \underline{\xi}, \Sigma^{-1} V \underline{\xi}) = (\underline{\xi}, V^t \Sigma^{-1} V \underline{\xi}) =$$
$$= (\underline{\xi}, -SV^{-1} S \Sigma^{-1} V \underline{\xi}) = (\underline{\xi}, -SV^{-1} \Omega V \underline{\xi}) = (\underline{\xi}, \Xi \underline{\xi}) = \qquad (22)$$
$$= \sum_{m=1}^{3} \frac{\xi_{2m-1}^2 + \xi_{2m}^2}{\varepsilon_m} = 2 \sum_{m=1}^{3} \frac{J_m}{\varepsilon_m}.$$

where we have introduced the normal mode action variables

$$J_m = \frac{\xi_{2m-1}^2 + \xi_{2m}^2}{2}. \qquad (23)$$

We can see that in the case of Gaussian distribution, $F = const \times \exp(-\Phi/2)$, parameters $\varepsilon_m$ really are the normal mode r.m.s. emittances.

Optics functions from the $\Sigma$-matrix.

Expression (8) for original variables via the normal form variables provides coupled optics functions:

$$\beta_{xm} = |(\underline{v}_{2m-1})_1|^2, \quad \beta_{ym} = |(\underline{v}_{2m-1})_3|^2, \quad \beta_{sm} = |(\underline{v}_{2m-1})_5|^2 \qquad (24)$$

From these so-called Mais-Ripken functions we can find the Edwards-Teng beta-functions as [1]

$$\beta_x = \beta_{x1} + \beta_{x2}, \quad \beta_y = \beta_{y1} + \beta_{y2}. \qquad (25)$$

For the projection of the 3$^{rd}$ (longitudinal) normal mode on the transverse planes we can introduce the dispersion functions using the recipe of Ref.[4]. Let the phase of the 3$^{rd}$ mode oscillation be such that its projection on longitudinal coordinate is zero $\zeta_5 \equiv s - v_0 t = V_{55}\xi_5 + V_{56}\xi_6 = 0$. Then all of the 3$^{rd}$ mode contribution to the transverse displacement will be due to the momentum offset $\delta \equiv \zeta_6 = V_{65}\xi_5 + V_{66}\xi_6 = (V_{66} - V_{65}V_{56}/V_{55})\xi_6$. Taking the ratio of the horizontal displacement $x \equiv \zeta_1 = V_{15}\xi_5 + V_{16}\xi_6 = (V_{16} - V_{15}V_{56}/V_{55})\xi_6$ to the momentum offset we recover the horizontal dispersion function and – analogously – the vertical one:

$$D_x = \frac{x}{\delta} = \frac{V_{16}V_{55} - V_{15}V_{56}}{V_{66}V_{55} - V_{65}V_{56}}, \quad D_y = \frac{y}{\delta} = \frac{V_{36}V_{55} - V_{35}V_{56}}{V_{66}V_{55} - V_{65}V_{56}}. \qquad (26)$$

Normal mode projections.

Another important question which requires knowledge of the transformation matrix V is establishing the correspondence between normal modes and planes in the phase space. To do this we can compare projections of a circle in the normalized space, $\xi_{2m-1} = \cos\varphi$, $\xi_{2m} = -\sin\varphi$, onto the planes of canonically conjugate variables. According to (7) $\zeta_{2p-1} = V_{2p-1, 2m-1}\xi_{2m-1} + V_{2p-1, 2m}\xi_{2m}$, $\zeta_{2p} = V_{2p, 2m-1}\xi_{2m-1} + V_{2p, 2m}\xi_{2m}$, where $p = 1, 2, 3$ is the plane number, and the normalized by $\pi$ area of the projected ellipse is

$$P(m \to p) = \frac{1}{\pi} \oint \zeta_{2p} d\zeta_{2p-1} =$$

$$= -\frac{1}{\pi} \int_0^{2\pi} (V_{2p,2m-1} \cos\varphi - V_{2p,2m} \sin\varphi)(V_{2p-1,2m-1} \sin\varphi + V_{2p-1,2m} \cos\varphi) d\varphi = \quad (27)$$

$$= V_{2p,2m} V_{2p-1,2m-1} - V_{2p,2m-1} V_{2p-1,2m} = (\underline{v}'_{2m-1})_{2p} (\underline{v}''_{2m-1})_{2p-1} - (\underline{v}''_{2m-1})_{2p} (\underline{v}'_{2m-1})_{2p-1}$$

Mechanical vs. canonical momenta.

We can use the described above method for computation of eigen-emittances to answer the often raised question of how important it is to use canonical momenta. Let us assume that we have a distribution of particles in axisymmetric magnetic field $B_z$ such that

$$<x^2> = <y^2> = \sigma^2, \quad <P_x^2> = <P_y^2> = \sigma_p^2, \quad (28)$$

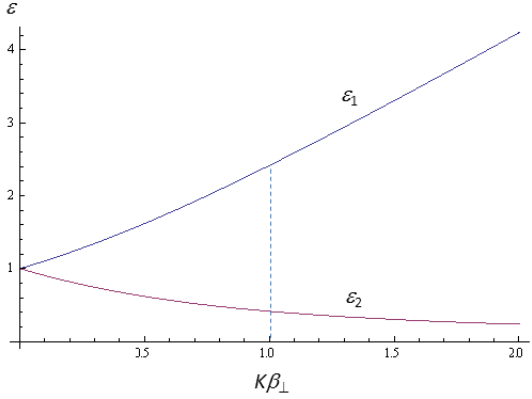

Figure 1. Eigen-emittances with mechanical momenta used.

with all other elements of the $\Sigma$-matrix being zero (we consider here a 4D case). Obviously both normal modes have equal emittances $\varepsilon_0 = \sigma_p \sigma$ and beta-functions $\beta_\perp = \sigma/\sigma_p$.

Now, if we would use the mechanical momenta, the non-zero covariance coefficients would be

$$<x^2> = <y^2> = \sigma^2, \quad <p_x^2> = <P_x^2> + K^2 <y^2> = \sigma_p^2 + K^2\sigma^2 = <p_y^2>,$$
$$<p_x y> = -<p_y x> = K\sigma^2, \quad K = B_z/2B\rho \quad (29)$$

Applying the developed method we would obtain for eigen-emittances

$$\varepsilon_{1,2}^2 = \varepsilon_0^2 (1 + 2K^2\beta_\perp^2 \pm 2|K|\beta_\perp \sqrt{1 + K^2\beta_\perp^2}) \quad (30)$$

Dependence of eigen-emittances on parameter $K\beta_\perp$ is shown in Fig.1. For the matched beta-function value $K\beta_\perp = 1$ and the eigen-emittances found with mechanical momenta differ by more than a factor of two from the correct value. However their product – the 4D emittance – remains correct: $\varepsilon_1\varepsilon_2 = \varepsilon_0^2$.

## 3. Suppression of the Halo Contribution

In a nonlinear system there are often halo particles which provide the largest contribution to the covariance matrix (4) thus distorting the value of emittances computed with it. Of course we can make cuts but since the emittances are not known in advance the procedure is ambiguous. The ambiguity can be eliminated if the halo suppression is made in a self-consistent way - most likely involving iterations - which takes into account the resulting $\Sigma$-matrix.

A natural idea – let us call it an "heuristic approach" – is to introduce weight depending on the particle position in the phase space and use recipe (5) for computing the averages. The exponential weight function,

$$w_k = \exp[-\frac{\alpha}{2}(\underline{\zeta}^{(k)}, \Sigma^{-1}\underline{\zeta}^{(k)})] \quad (31)$$

with $\alpha$ being a free parameter, can provide an efficient rejection of the halo particles. For the $\Sigma$-matrix we then have the following equation (in absence of other weights)

$$\Sigma_{ij} = \sum_{k=1}^{N} \zeta_i^{(k)} \zeta_j^{(k)} \exp[-\frac{\alpha}{2}(\underline{\zeta}^{(k)}, \Sigma^{-1} \underline{\zeta}^{(k)})] \bigg/ \sum_{k=1}^{N} \exp[-\frac{\alpha}{2}(\underline{\zeta}^{(k)}, \Sigma^{-1} \underline{\zeta}^{(k)})] \qquad (32)$$

which can be solved iteratively.

However, besides suppression of the halo contribution, an acceptable method must provide the exact result in absence of the halo. Figures 2 and 3 show the results of tests of the heuristic method in the 1D case. First, $\Sigma = \sigma^2$ was computed by this method for 25 realizations (seeds) of the Gaussian distribution with $\sigma = 1$ for different number of particles $N$ and three values of parameter $\alpha$, $\alpha = 0$ providing the conventional value as in eq.(4).

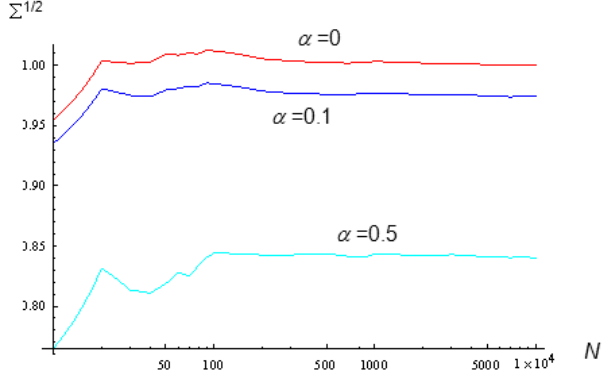
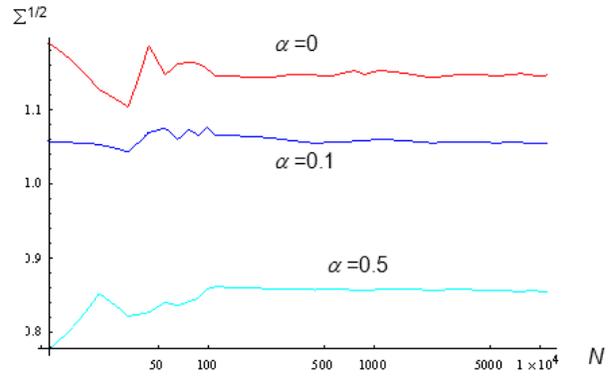

Figure 2. Square root of $\Sigma$ from eq.(32) averaged over 25 realizations of 1D Gaussian distribution with $\sigma = 1$ as function of the number of particles $N$.

Figure 3. Square root of $\Sigma$ from eq.(32) averaged over 25 realizations of superposition of 1D Gaussian distributions with $\sigma = 1 (90\%)$ and $\sigma = 3 (10\%)$.

Then the heuristic approach was applied to a superposition of two 1D Gaussian distributions with $\sigma = 1 (90\%)$ and $\sigma = 3 (10\%)$ representing the beam core and a halo. As we can see the value $\alpha = 0.1$ is still insufficient to suppress the halo contribution (Fig. 3) but already eats into the beam core (Fig. 2).

We can conclude that the heuristic approach has an inherent ambiguity and does not provide accurate results.

## 4. Nonlinear Fit of the Klimontovich Distribution

A more rigorous approach is based on a nonlinear fit of particle distribution in the phase space known from measurements or simulations

$$G(\underline{z}) = \frac{1}{N} \sum_{k=1}^{N} \delta_{6D}(\underline{z} - \underline{z}^{(k)}) \equiv \frac{1}{N} \sum_{k=1}^{N} \prod_{i=1}^{6} \delta(z_i - z_i^{(k)}) \qquad (33)$$

where $\delta_{6D}$ is six-dimensional Dirac's $\delta$-function.

Distribution (33) is sometimes referred to as the Klimontovich distribution. Our task is to approximate it with a smooth function. We will employ Gaussian distribution, though other functions (e.g. parabolic) can be used:

$$F(\underline{\zeta}) = \frac{\eta}{(2\pi)^{n/2} \sqrt{\det \Sigma}} \exp[-\frac{1}{2}(\underline{\zeta}, \Sigma^{-1} \underline{\zeta})]. \qquad (34)$$

The fitting parameters are the elements of the $\Sigma$-matrix and – optionally – the mean values of coordinates and momenta and parameter $\eta$ which can be considered as the fraction of particles in the beam core.

For a moment let us replace point-like particles in distribution *G* with spheres of radius $\rho$ so that *G* was integrable with square and later put $\rho \to 0$. Keeping $\rho$ finite, the fitting can be formulated as a minimization problem,

$$\int_{-\infty}^{\infty}\!\!\!...\int_{-\infty}^{\infty} |F-G|^2 \, dz_1...dz_n = \int_{-\infty}^{\infty}\!\!\!...\int_{-\infty}^{\infty} (F^2 - 2FG) \, dz_1...dz_n + \int_{-\infty}^{\infty}\!\!\!...\int_{-\infty}^{\infty} G^2 \, dz_1...dz_n \to \min \quad (35)$$

Since the last integral does not depend on the fitting parameters the problem can be reformulated as a search for the maximum of the first term taken with the opposite sign. Since G enters this term linearly we can put $\rho \to 0$ and in the case of Gaussian distribution come to the maximization problem

$$\int_{-\infty}^{\infty}\!\!\!...\int_{-\infty}^{\infty} (2FG - F^2) \, dz_1...dz_n = \frac{\eta}{(2\pi)^{n/2}\sqrt{\det \Sigma}} \left\{ \frac{2}{N} \sum_{k=1}^{N} \exp[-\frac{1}{2}(\underline{\zeta}^{(k)}, \Sigma^{-1} \underline{\zeta}^{(k)})] - \frac{\eta}{2^{n/2}} \right\} \to \max \quad (36)$$

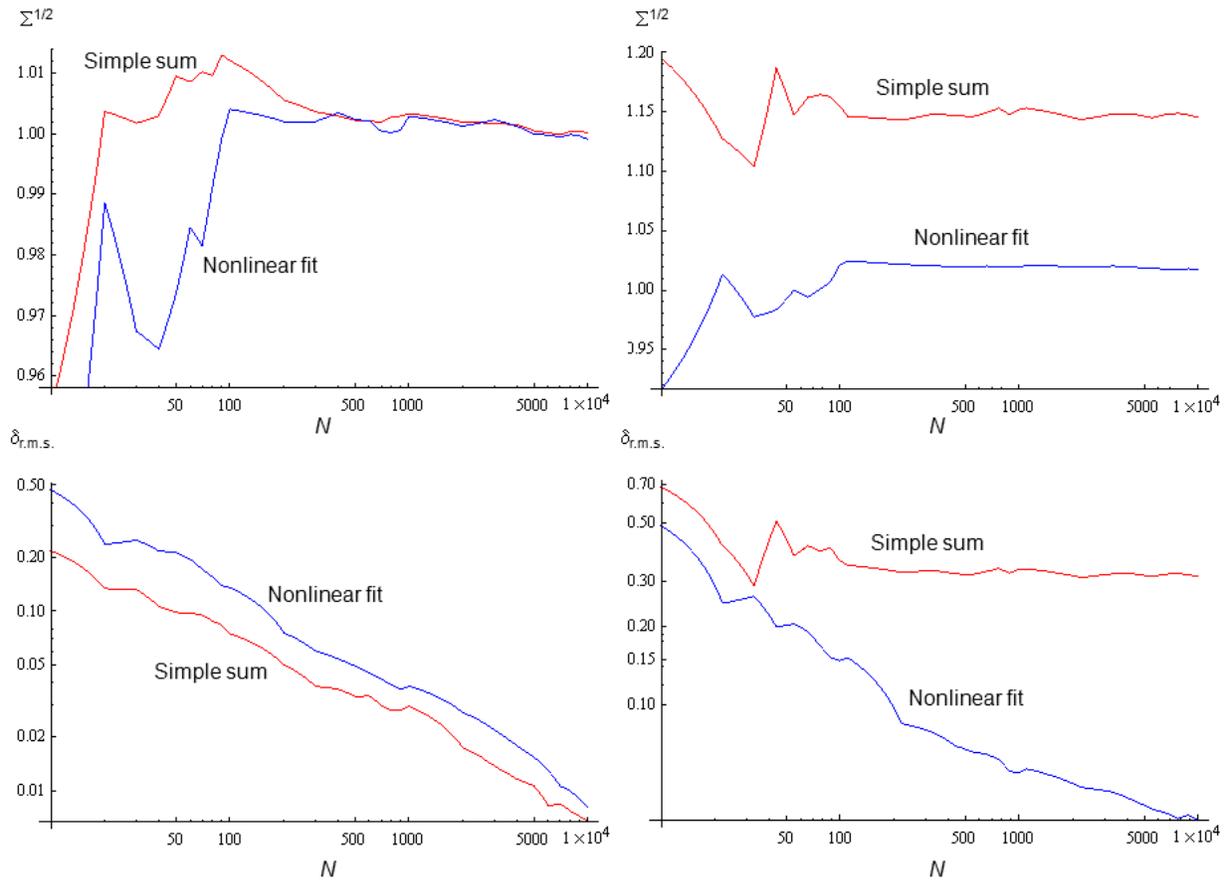

Figure 4. Square root of Σ (top row) and its r.m.s. error (bottom row) obtained by nonlinear fit with variable parameter $\eta$ for 25 realizations of 1D Gaussian distribution with $\sigma = 1$ (left) and of superposition of two such distributions with $\sigma = 1$ (90%) and $\sigma = 3$ (10%) (right).

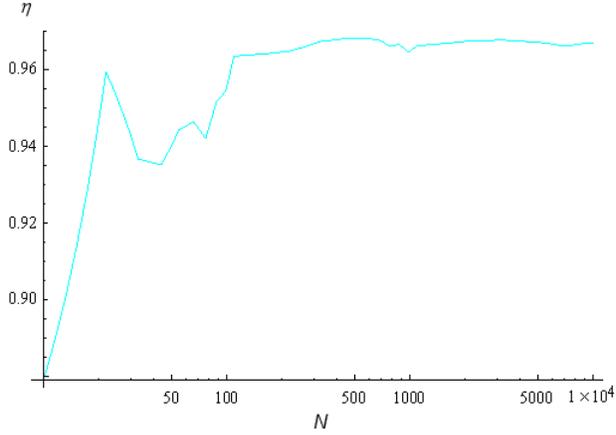

Figure 5. Average value of parameter $\eta$ obtained by the fit for the superposition of Gaussian distributions corresponding to the right column of Fig. 4

Figure 4 shows the result of 1D tests obtained with the *Mathematica* NMaximize command under the same conditions as in Figs. 2 and 3. By "simple sum" the $\Sigma$-value is meant which was computed according to eq.(4) and coincides with case $\alpha=0$ in Figs. 2, 3. For a Gaussian distribution the fit gives very close to $\sigma=1$ value though with somewhat larger variance than the simple sum.

For the superposition of two Gaussian distributions the fit effectively rejects the halo providing square root of average value $\Sigma^{1/2}=0.997$ for $N=10^4$ particles. The fraction of particles in the beam core according to the fit is $\eta=0.967$ so that two thirds of the halo particles were absorbed into the core.

In a 1D case there were only 3 fitting parameters – mean value, $\Sigma$ and $\eta$ – so the maximization process was converging quite fast. Generally there is $n(n+3)/2+1$ fitting parameters: $n(n+1)/2$ independent elements of the $\Sigma$-matrix, $n$ mean values of coordinates and $\eta$. For $n=6$ this makes total of 28 parameters and the direct maximization process occurs too slow to be practical.

## 5. Iterative Procedure

To obtain equations suitable for iterations we can take advantage of the fact that at the maximum of the merit function (36) which we will rewrite dropping numerical coefficient

$$H(\Sigma,\underline{a},\eta) = \frac{\eta}{\sqrt{\det\Sigma}}\left\{\frac{1}{N}\sum_{k=1}^{N}\exp[-\frac{1}{2}(\underline{\zeta}^{(k)},\Sigma^{-1}\underline{\zeta}^{(k)})] - \frac{\eta}{2^{n/2+1}}\right\}, \quad \underline{\zeta}^{(k)} = \underline{z}^{(k)} - \underline{a}, \tag{37}$$

all its derivatives w.r.t. fitting parameters turn zero.

Let us first consider $H$ derivative by $(\Sigma^{-1})_{ji}$ taking notice that $\det\Sigma=1/\det(\Sigma^{-1})$ and making use of Jacobi's formula for a derivative of a determinant. According to the latter

$$\frac{d}{dA_{ji}}\det A = \det A \cdot (A^{-1})_{ij} \tag{38}$$

so that

$$\frac{d}{d(\Sigma^{-1})_{ji}}\frac{1}{\sqrt{\det\Sigma}} = \frac{d}{d(\Sigma^{-1})_{ji}}\sqrt{\det\Sigma^{-1}} = \frac{1}{2}\sqrt{\det\Sigma^{-1}}\cdot\Sigma_{ij} = \frac{1}{2\sqrt{\det\Sigma}}\Sigma_{ij} \tag{39}$$

In the result we have:

$$\frac{d}{d(\Sigma^{-1})_{ji}}H(\Sigma,\underline{a},\eta) = 0 =$$

$$-\frac{\eta}{2\sqrt{\det\Sigma}}\frac{1}{N}\sum_{k=1}^{N}\zeta_i^{(k)}\zeta_j^{(k)}\exp[-\frac{1}{2}(\underline{\zeta}^{(k)},\Sigma^{-1}\underline{\zeta}^{(k)})] + \frac{1}{2}H(\Sigma,\underline{a},\eta)\cdot\Sigma_{ij} \tag{40}$$

or

$$\Sigma_{ij} = \frac{1}{N}\sum_{k=1}^{N}\zeta_i^{(k)}\zeta_j^{(k)}\exp[-\frac{1}{2}(\underline{\zeta}^{(k)},\Sigma^{-1}\underline{\zeta}^{(k)})] \bigg/ \left\{\frac{1}{N}\sum_{k=1}^{N}\exp[-\frac{1}{2}(\underline{\zeta}^{(k)},\Sigma^{-1}\underline{\zeta}^{(k)})] - \frac{\eta}{2^{n/2+1}}\right\} \tag{41}$$

This equation for the Σ-matrix elements cannot be obtained by introducing weights (unless $\eta=0$). The fitted value of $\eta$ can be readily found from condition $dH/d\eta=0$ and amounts to

$$\eta = \frac{2^{n/2}}{N}\sum_{k=1}^{N}\exp[-\frac{1}{2}(\underline{\zeta}^{(k)}, \Sigma^{-1}\underline{\zeta}^{(k)})] \qquad (42)$$

With this value of $\eta$ the equation for the Σ-matrix elements becomes

$$\Sigma_{ij} = 2\sum_{k=1}^{N}\zeta_i^{(k)}\zeta_j^{(k)}\exp[-\frac{1}{2}(\underline{\zeta}^{(k)}, \Sigma^{-1}\underline{\zeta}^{(k)})] \Big/ \sum_{k=1}^{N}\exp[-\frac{1}{2}(\underline{\zeta}^{(k)}, \Sigma^{-1}\underline{\zeta}^{(k)})] \qquad (43)$$

It contains an extra factor of 2 (!) compared to eq.(32) of the heuristic approach with $\alpha=1$.

Finally let us consider $H$ derivative w.r.t. the mean values of coordinates which enter $H$ via $\underline{\zeta}^{(k)}$ as shown in eq.(37). Taking into account symmetry of the Σ-matrix and its inverse we get

$$\begin{aligned}\frac{d}{d\underline{a}}H(\Sigma, \underline{a}, \eta) = 0 &= \frac{\eta}{2\sqrt{\det\Sigma}}[\Sigma^{-1}+(\Sigma^{-1})^t]\frac{1}{N}\sum_{k=1}^{N}\underline{\zeta}^{(k)}\exp[-\frac{1}{2}(\underline{\zeta}^{(k)}, \Sigma^{-1}\underline{\zeta}^{(k)})] \\ &= \frac{\eta}{\sqrt{\det\Sigma}}\Sigma^{-1}\frac{1}{N}\sum_{k=1}^{N}(\underline{z}^{(k)}-\underline{a})\exp[-\frac{1}{2}(\underline{\zeta}^{(k)}, \Sigma^{-1}\underline{\zeta}^{(k)})]\end{aligned} \qquad (44)$$

Since the inverse Σ-matrix does not have a null space, the multiplying vector should be zero, hence

$$\underline{a} = \sum_{k=1}^{N}\underline{z}^{(k)}\exp[-\frac{1}{2}(\underline{\zeta}^{(k)}, \Sigma^{-1}\underline{\zeta}^{(k)})] \Big/ \sum_{k=1}^{N}\exp[-\frac{1}{2}(\underline{\zeta}^{(k)}, \Sigma^{-1}\underline{\zeta}^{(k)})] \qquad (45)$$

This equation can be obtained with weight (31) with $\alpha=1$, but this is the only commonality with the heuristic approach.

Equations (41), (42) and (45) can be solved together iteratively and in the 1D case provide exactly the same results as the direct maximization process with merit function (36) or (37). In the 6D case it takes 20-30 iterations with 28 fitting parameters to achieve precision $10^{-6}$ for $N=10^4$ particles. *Mathematica* spends on this about 12s, FORTRAN or C code can do this by orders of magnitude faster.

In real applications it may not be necessary to fit all parameters. For example we can fix $\eta=1$ to have a better representation of the whole ensemble (still there will be drastic suppression of the tails).

For values of $\eta$ close to 1 some damping should be introduced to avoid oscillations around the solution. The Σ-matrix at iteration $l$ is then presented as

$$\Sigma^{(l)} = (1-d)\Sigma^{(l-1)} + d\cdot\Sigma^{(\text{formula})} \quad , \qquad (46)$$

where $\Sigma^{(\text{formula})}$ is the value given by eq.(41) and $d$ is the damping factor (for $\eta=1$ $d\approx 0.8$). With $\eta$ included in the iterative process the damping is not necessary ($d=1$).

## 6. Application of the algorithm for 6D ionization cooling in HFOFO channel

Let us apply the proposed algorithm for 6D cooling analysis in a Helical FOFO channel [5] which can be used to cool both $\mu^+$ and $\mu^-$ beams simultaneously.

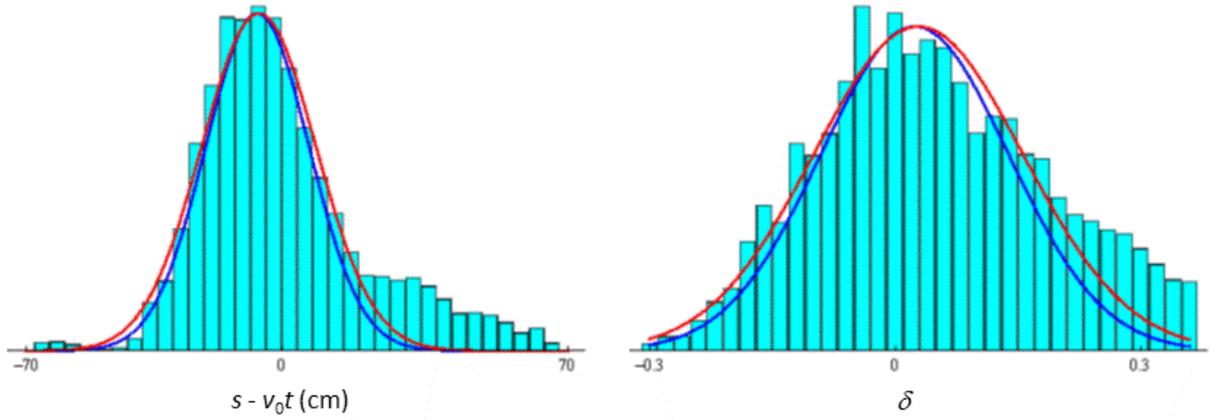

Figure 6. Projections onto the longitudinal coordinate (left) and $\delta$ (right) of the original particle distribution (cyan bars) and of its Gaussian fit with $\eta=1$ and $\eta=\eta_{fit}$ (red and blue solid lines respectively).

First we apply the algorithm to the $\mu^+$ beam from the front end consisting of buncher and rotator (some early version provided by C. Yoshikawa was used). Figure 6 shows histograms of particle distribution in the longitudinal coordinate and $\delta$ (see eqs. (1) and (3) for definitions), as well as the projections of fitted distributions with parameter $\eta$ fixed ($\eta=1$) and fitted in the process ($\eta=\eta_{fit}$). The solid lines were slightly shifted by the same amount for easier comparison of the distribution widths.

When computing a projection of the particle distribution onto one axis in the phase space the distribution must be integrated over all other coordinates. Using eq. (A.1) and performing some algebraic manipulations it is possible to show that the projection of the Gaussian distribution onto the $m^{th}$ axis is proportional to

$$\exp\left\{-\frac{1}{2}(\Sigma^{-1})_{mm}\zeta_m^2[2-(\Sigma^{-1})_{mm}\Sigma_{mm}]\right\}. \tag{47}$$

If the $\Sigma$-matrix is diagonal then $(\Sigma^{-1})_{mm}=1/\Sigma_{mm}$ and eq. (47) is reduced to the conventional formula.

As Fig. 6 shows, the fit with $\eta=1$ provides almost as strong halo suppression as with $\eta$ included in the fitting. The corresponding values of all three eigen-emittances are given in Table 1.

| $\eta$ | $\varepsilon_{1N}$ (cm) | $\varepsilon_{2N}$ (cm) | $\varepsilon_{3N}$ (cm) |
|---|---|---|---|
| 1 | 1.59 | 1.42 | 3.94 |
| $\eta_{fit}=0.67$ | 1.26 | 1.15 | 3.20 |

Table 1. Eigen-emittances computed with fixed ($\eta=1$) and fitted ($\eta=\eta_{fit}$) parameter $\eta$.

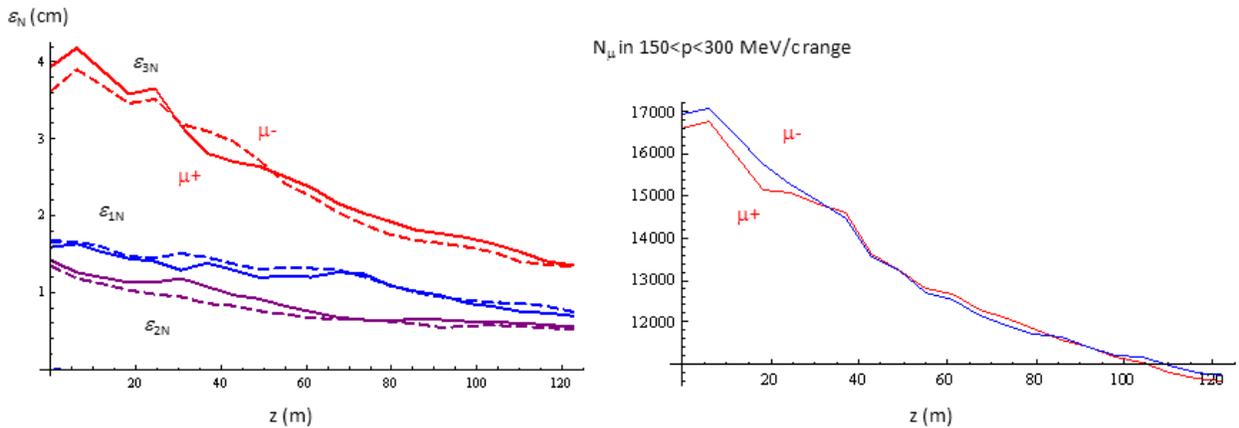

Figure 7. Evolution of the eigen-emittances of the μ⁺ and μ⁻ beams in HFOFO channel computed with $\eta =1$ (left) and number of muons in the momentum range 150< $p$ <300 MeV/c (right).

|  | $\varepsilon_{1N}$ (cm) | $\varepsilon_{2N}$ (cm) | $\varepsilon_{3N}$ (cm) |
|---|---|---|---|
| initial | 1.59 | 1.42 | 3.94 |
| final | 0.70 | 0.56 | 1.36 |
| ini / fin | 2.27 | 2.56 | 2.91 |

Table 2. The initial and final values of the eigen-emittances and their ratio.

The performance of the channel is illustrated by Fig. 7 and Table 2. The 6D cooling factor is $\varepsilon_{6D}^{(ini)}/\varepsilon_{6D}^{(fin)} = 16.9$.

## Appendix. Useful multi-dimensional integrals

In the course of this work some useful mathematical formulae was obtained which cannot be found (or are incorrect) in handbooks. We present two multidimensional integrals here.

The basic integral with a positive-definite matrix A and *n*-dimensional vector $\underline{b}$:

$$\int_{-\infty}^{\infty}...\int_{-\infty}^{\infty} e^{-(\underline{x},A\underline{x})+2(\underline{b},\underline{x})} dx_1...dx_n = \frac{\pi^{n/2}}{\sqrt{\det A}} e^{(\underline{b},A^{-1}\underline{b})} \qquad (A.1)$$

The other integral can be used in the 2ⁿᵈ order moments calculation:

$$\int_{-\infty}^{\infty}...\int_{-\infty}^{\infty} (\underline{x},C\underline{x}) e^{-(\underline{x},A\underline{x})} dx_1...dx_n = \frac{\pi^{n/2}}{2\sqrt{\det A}} \text{Tr}(A^{-1}C) \qquad (A.2)$$

where C is arbitrary *n*×*n* matrix.